\documentclass[conference]{IEEEtran-v17}
\usepackage{xspace}
\usepackage{url}
\usepackage[cmex10]{amsmath}
\usepackage{bbm}
\usepackage[dvips]{graphicx}
  \DeclareGraphicsExtensions{.eps,.pdf}
\usepackage{paralist}

\usepackage[dvipdfm]{hyperref}
\usepackage{fancyref} 
\usepackage{amsmath}
\usepackage{amssymb}
\usepackage{footnote}

\usepackage[nosort]{cite}
\usepackage{color}

\usepackage{psfrag}

\usepackage{balance}

\usepackage{bm}
\usepackage{upgreek}
\usepackage{amssymb}
\usepackage{amsfonts}
\usepackage{mathrsfs}
\usepackage{xspace}

\newcommand{\safemath}[2]{\newcommand{#1}{\ensuremath{#2}\xspace}}
\safemath{\opE}{\mathbb{E}}
\newcommand{\Ex}[2]{\ensuremath{\opE_{#1}\lefto[#2\right]}} 	
\newcommand{\lefto}{\mathopen{}\left}

\safemath{\indist}{\mathsf{P}}
\safemath{\outdist}{\mathsf{Q}}
\safemath{\inpdf}{\mathsf{p}}
\safemath{\outpdf}{\mathsf{q}}

\safemath{\blocklength}{L}
\safemath{\codelength}{n}
\safemath{\cohtime}{T}
\safemath{\blockindex}{l}
\safemath{\NumCode}{M}
\safemath{\error}{\epsilon}
\safemath{\RXant}{M}
\safemath{\snr}{\rho}
\safemath{\mi}{I}
\safemath{\difent}{\mathrm{h}}		
\newcommand{\given}{\,\vert\,}				
\safemath{\define}{\triangleq}			
\safemath{\fnorm}{\mathrm{F}}
\safemath{\rankcorr}{Q}
\safemath{\sqrtmat}{\mathbf{P}}

\safemath{\veca}{\mathbf{a}}
\safemath{\vecb}{\mathbf{b}}
\safemath{\vecc}{\mathbf{c}}
\safemath{\vecd}{\mathbf{d}}
\safemath{\vece}{\mathbf{e}}
\safemath{\vecf}{\mathbf{f}}
\safemath{\vecg}{\mathbf{g}}
\safemath{\vech}{\mathbf{h}}
\safemath{\veci}{\mathbf{i}}
\safemath{\vecj}{\mathbf{j}}
\safemath{\veck}{\mathbf{k}}
\safemath{\vecl}{\mathbf{l}}
\safemath{\vecm}{\mathbf{m}}
\safemath{\vecn}{\mathbf{n}}
\safemath{\veco}{\mathbf{o}}
\safemath{\vecp}{\mathbf{p}}
\safemath{\vecq}{\mathbf{q}}
\safemath{\vecr}{\mathbf{r}}
\safemath{\vecs}{\mathbf{s}}
\safemath{\vect}{\mathbf{t}}
\safemath{\vecu}{\mathbf{u}}
\safemath{\vecv}{\mathbf{v}}
\safemath{\vecw}{\mathbf{w}}
\safemath{\vecx}{\mathbf{x}}
\safemath{\vecy}{\mathbf{y}}
\safemath{\vecz}{\mathbf{z}}

\safemath{\matA}{\mathbf{A}}
\safemath{\matB}{\mathbf{B}}
\safemath{\matC}{\mathbf{C}}
\safemath{\matD}{\mathbf{D}}
\safemath{\matE}{\mathbf{E}}
\safemath{\matF}{\mathbf{F}}
\safemath{\matG}{\mathbf{G}}
\safemath{\matH}{\mathbf{H}}
\safemath{\matI}{\mathbf{I}}
\safemath{\matJ}{\mathbf{J}}
\safemath{\matK}{\mathbf{K}}
\safemath{\matL}{\mathbf{L}}
\safemath{\matM}{\mathbf{M}}
\safemath{\matN}{\mathbf{N}}
\safemath{\matO}{\mathbf{O}}
\safemath{\matP}{\mathbf{P}}
\safemath{\matQ}{\mathbf{Q}}
\safemath{\matR}{\mathbf{R}}
\safemath{\matS}{\mathbf{S}}
\safemath{\matT}{\mathbf{T}}
\safemath{\matU}{\mathbf{U}}
\safemath{\matV}{\mathbf{V}}
\safemath{\matW}{\mathbf{W}}
\safemath{\matX}{\mathbf{X}}
\safemath{\matY}{\mathbf{Y}}
\safemath{\matZ}{\mathbf{Z}}

\safemath{\matSigma}{\mathbf{\Sigma}}

\safemath{\diag}{\mathrm{diag}}

\safemath{\jpg}{\mathcal{CN}}			
\safemath{\complexset}{\mathbb{C}}

\newcommand{\tp}[1]{\ensuremath{#1^{T}}} 		
\newcommand{\herm}[1]{\ensuremath{#1^{H}}} 	

\safemath{\veczero}{\mathbf{0}} 

\safemath{\coh}{\mathrm{coh}}
\safemath{\bigo}{\mathcal{O}}
\begin{document}
\IEEEoverridecommandlockouts

\title{Diversity  versus Channel Knowledge \\at Finite Block-Length \thanks{Tobias Koch has received funding from the
European Community's Seventh Framework Programme (FP7/2007-2013) under grant agreement No. 252663.}}

\author{\IEEEauthorblockN{Wei Yang$^1$, Giuseppe Durisi$^1$, Tobias Koch$^2$, and Yury Polyanskiy$^3$}
\\
\IEEEauthorblockA{
$^1$Chalmers University of Technology, 41296 Gothenburg, Sweden\\
$^2$Universidad Carlos III de Madrid, 28911 Legan\'{e}s, Spain\\
$^3$Massachusetts Institute of Technology, Cambridge, MA, 02139 USA
}}

\maketitle

\begin{abstract}
We study the maximal achievable rate $R^\ast(\codelength, \error)$ for a given block-length $\codelength$ and block error probability $\error$ over Rayleigh block-fading channels in the noncoherent setting and in the finite block-length regime. Our results show that for a given block-length and error probability, $R^\ast(\codelength, \error)$ is not monotonic in the channel's coherence time, but there exists a rate maximizing coherence time that optimally trades between diversity and cost of estimating the channel.
\end{abstract}

\section{Introduction}
It is well known that the capacity of the single-antenna Rayleigh-fading channel with perfect channel state information (CSI) at the receiver (the so-called \emph{coherent setting}) is independent of the fading dynamics~\cite{biglieri98-10a}. In practical wireless systems, however, the channel is usually not known \emph{a priori} at the receiver and must be estimated, for example, by transmitting training symbols. An important observation is that the training overhead is a function of the channel dynamics, because the faster the channel varies, the more training symbols are needed in order to estimate the channel accurately~\cite{lapidoth02-05a,hassibi03,vikalo04-09}. One way to determine the training overhead, or more generally, the capacity penalty due to lack of channel knowledge, is to study capacity in the \emph{noncoherent setting}, where neither the transmitter nor the receiver are assumed to have \emph{a priori} knowledge of the realizations of the fading channel (but both are assumed to know its statistics perfectly)~\cite{lapidoth05-02a}.

In this paper, we model the fading dynamics using the well-known block-fading model~\cite{marzetta99-01a,hochwald00-03a,Zheng02-02a} according to which the channel coefficients remain constant for a period of $\cohtime$ symbols, and change to a new independent realization in the next period. The parameter $\cohtime$ can be thought of as the channel's coherence time. Unfortunately, even for this simple model, no closed-form expression for capacity is available to date. A capacity lower bound based on the \emph{isotropically distributed (i.d.)} unitary distribution is reported in \cite{marzetta99-01a}. In \cite{hochwald00-03a,Zheng02-02a,durisi11-08a}, it is shown that capacity in the high signal-to-noise ratio (SNR) regime grows logarithmically with SNR, with the \emph{pre-log} (defined as the asymptotic ratio between capacity and the logarithm of SNR as SNR goes to infinity) being $1-1/T$. This agrees with the intuition that the capacity penalty due to lack of a priori channel knowledge at the receiver is small when the channel's coherence time is large.

In order to approach capacity, the block-length $\codelength$ of the codewords must be long enough to average out the fading effects (i.e., $\codelength \gg \cohtime$). Under practical delay constraints, however, the actual performance metric is the maximal achievable rate $R^\ast(\codelength, \error)$ for a given block-length $\codelength$ and block error probability $\error$. By studying $R^\ast(\codelength, \error)$ for the case of fading channels and in the coherent setting, Polyanskiy and Verd\'{u} recently showed that faster fading dynamics are advantageous in the finite block-length regime when the channel is known to the receiver~\cite{polyanskiy-11-isit}, because faster fading dynamics yield larger diversity gain.

We expect that the maximal achievable rate $R^\ast(\codelength, \error)$ over fading channels in the \emph{noncoherent setting} and in the \emph{finite block-length regime} is governed by two effects working in opposite directions: when the channel's coherence time decreases, we can code the information over a larger number of independent channel realizations, which provides higher diversity gain, but we need to transmit training symbols more frequently to learn the channel accurately, which gives rise to a rate loss.

In this paper, we shed light on this fundamental tension by providing upper and lower bounds on $R^\ast(\codelength, \error)$ in the noncoherent setting.
For a given block-length and error probability, our bounds show that there exists  indeed a rate-maximizing channel's coherence time that optimally trades between diversity and cost of estimating the channel.
%


\paragraph*{Notation} 
\label{sec:notation}
Uppercase boldface letters denote matrices and lowercase boldface letters designate vectors. Uppercase sans-serif letters (e.g.,~$\mathsf{Q}$) denote probability distributions, while lowercase sans-serif letters  (e.g.,~$\mathsf{r}$) are reserved for probability density functions (pdf).
The superscripts~$\tp{}$ and~$\herm{}$ stand for transposition and Hermitian transposition, respectively. We denote the identity matrix of dimension $T\times T$ by $\matI_{T}$; the sequence of vectors $\{\veca_1,\ldots,\veca_n\}$ is written as $\veca^n$. We denote expectation and variance by~$\Ex{}{\cdot}$ and~$\mathrm{Var}[\cdot]$, respectively, and use the notation~$\Ex{\vecx}{\cdot}$ or $\Ex{\indist_{\vecx}}{\cdot}$ to stress that expectation is taken with respect to $\vecx$ with distribution~$\indist_\vecx$. The relative entropy between two distributions $\indist$ and $\outdist$ is denoted by $D(\indist\|\outdist)$ \cite[Sec.~8.5]{cover06-a}. For two functions~$f(x)$ and~$g(x)$, the
notation~$f(x) = \bigo(g(x))$, $x\to \infty$, means that
$\lim \sup_{x\to\infty}\bigl|f(x)/g(x)\bigr|<\infty$, and
$f(x) = o(g(x))$, $x\to \infty$, means that $\lim_{x\to\infty}\bigl|f(x)/g(x)\bigr|=0$.
Furthermore, $\jpg(\veczero,\matR)$ stands for the distribution of a circularly-symmetric complex Gaussian random vector with covariance matrix $\matR$, and $\mathrm{Gamma}(\alpha,\beta)$ denotes the gamma distribution~\cite[Ch.~17]{johnson95-1} with parameters $\alpha$ and $\beta$. Finally, $\log(\cdot)$ indicates the natural logarithm, $\Gamma(\cdot)$ denotes the gamma function \cite[Eq.~(6.1.1)]{abramowitz72}, and $\psi(\cdot)$ designates the digamma function~\cite[Eq.~(6.3.2)]{abramowitz72}.

\section{Channel Model and Fundamental Limits}
We consider a single-antenna Rayleigh block-fading channel with \emph{coherence time} $\cohtime$. Within the $l$th coherence interval, the channel input-output relation can be written as
\begin{equation}
\label{eq:channel_IO_finitebl}
\vecy_l = s_l\vecx_l +\vecw_l
\end{equation}
where $\vecx_l$ and $\vecy_l$ are the input and output signals, respectively, $\vecw_l \sim \jpg(0, \matI_\cohtime)$ is the additive noise, and $s_l\sim\jpg(0,1)$ models the fading, whose realization we assume is not known at the transmitter and receiver (noncoherent setting). In addition, we assume that  $\{s_l\}$ and $\{\vecw_l\}$ take on independent realizations over successive coherence intervals.

We consider channel coding schemes employing codewords of length $\codelength=\blocklength \cohtime$. Therefore, each codeword spans $\blocklength$ independent fading realizations. Furthermore, the codewords are assumed to satisfy the following power constraint
\begin{align}
\label{eq:codeword-power-constraint}
\sum\limits_{l=1}^{\blocklength} \|\vecx_l\|^2 \leq \blocklength\cohtime\snr .
\end{align}
Since the variance of $s_l$ and of the entries of $\vecw_l$ is normalized to one, $\snr$ in (\ref{eq:codeword-power-constraint}) can be interpreted as the SNR at the receiver.

Let $R^\ast(\codelength,\error)$ be the maximal achievable rate among all codes with block-length $\codelength$ and decodable with probability of error $\error$. For every fixed $\cohtime$ and $\error$, we have\footnote{The subscript $l$ is omitted whenever immaterial.}
\begin{align}
\label{eq:def-capacity}
\lim\limits_{\codelength \to \infty} R^\ast(\codelength,\error) = C(\snr) = \frac{1}{\cohtime} \sup_{\indist_{\vecx}} I(\vecx;\vecy)
\end{align}
where $C(\snr)$ is the capacity of the channel in~(\ref{eq:channel_IO_finitebl}), $I(\vecx;\vecy)$ denotes the mutual information between $\vecx$ and $\vecy$, and the supremum in (\ref{eq:def-capacity}) is taken over all input distributions $\indist_{\vecx}$ that satisfy
 \begin{equation}
\label{eq:average-power-constraint}
\Ex{}{\|\vecx\|^2} \leq \cohtime\snr.
\end{equation}

No closed-form expression of $C(\snr)$ is available to date. The following lower bound $L(\snr)$ on $C(\snr)$ is reported in~\cite[Eq.~(12)]{marzetta99-01a}
\begin{align}
\label{eq:lower-bound}
L(\snr) &= \frac{1}{\cohtime}\left((\cohtime-1)\log\lefto(\cohtime\snr\right)-\log\Gamma(\cohtime)-\cohtime +\frac{\cohtime(1+\snr)}{1+\cohtime\snr}\right)\notag\\
&\quad - \frac{1}{\cohtime}\int\nolimits_{0}^{\infty}e^{-u}
 \tilde{\gamma}\left(\cohtime-1, \cohtime\snr u\right) \left(1+ \frac{1}{\cohtime\snr}\right)^{\cohtime-1}\notag\\
&\quad\quad \quad\quad\quad\quad\quad\quad \times \log\lefto(u^{1-\cohtime}\tilde{\gamma}\lefto(\cohtime-1,\cohtime\snr u\right)\right)du
\end{align}
where
\begin{equation*}
\tilde{\gamma}(n,x) \define\frac{1}{\Gamma(n)} \int \nolimits_{0}^{x} t^{n-1}e^{-t} dt
\end{equation*}
denotes the \emph{regularized incomplete gamma function}. The input distribution used in \cite{marzetta99-01a} to establish (\ref{eq:lower-bound}) is the i.d. unitary distribution, where the input vector takes on the form $\vecx = \sqrt{\cohtime\snr}\, \vecu_\vecx$ with $\vecu_\vecx$ uniformly distributed on the unit sphere in $\complexset^{\cohtime}$. We shall denote this input distribution as $\indist^{(\mathrm{U})}_{\vecx}$. It can be shown that $L(\snr)$ is asymptotically tight at high SNR (see \cite[Thm.~4]{hochwald00-03a}), i.e.,
\begin{equation*}
C(\snr) = L(\snr) + o(1), \quad\quad\snr \rightarrow \infty.
\end{equation*}

\section{Bounds on $R^\ast(\codelength, \error )$}

\subsection{Perfect-Channel-Knowledge Upper Bound}
\label{sec:perfect-CSI-upperbound}
We establish a simple upper bound on $R^\ast(\codelength, \error )$ by assuming that the receiver has perfect knowledge of the realizations of the fading process $\{s_{l}\}$.
Specifically, we have that
\begin{align}
R^\ast(\codelength, \error ) \leq R_{\coh}^\ast(\codelength, \error )
\end{align}
where $R_{\coh}^\ast(\codelength, \error )$ denotes the maximal achievable rate for a given block-length $\codelength$ and probability of error $\error$ in the coherent setting.

By generalizing the method used in~\cite{polyanskiy-11-isit} for stationary ergodic fading channels to the present case of block-fading channels, we obtain the following asymptotic expression for~$R_{\coh}^\ast(\codelength, \error )$:
\begin{IEEEeqnarray}{rCl}
\label{eq:achieve-rate-coh}
R_{\coh}^\ast(\codelength,\error) &=& C_{\coh}(\snr) - \sqrt{\frac{V_{\coh}(\snr)}{\codelength}}Q^{-1}(\error)\notag\\
&&+\: o\lefto(\frac{1}{\sqrt{n}}\right), \quad n\to \infty.
\end{IEEEeqnarray}
Here,  $C_{\mathrm{coh}}(\snr)$ is the capacity of the block-fading channel in the coherent setting, which is given by~\cite[Eq.~(3.3.10)]{biglieri98-10a}
\begin{align}
 \label{eq:capacity-coherent}
 C_{\mathrm{coh}}(\snr) = \Ex{s}{\log\lefto(1+|s|^2\snr\right)}
 \end{align}
 $Q(x)=\int\nolimits_{x}^{\infty}\frac{1}{\sqrt{2\pi}}e^{-t^2/2} dt$ denotes the $Q$-function,
and
\begin{align*}
V_\coh(\snr)  =  \cohtime \mathrm{Var}\lefto[\log\lefto(1+ \snr |s|^2\right)\right] + 1 - \mathbb{E}^2\lefto[\frac{1}{1+ \snr |s|^2}\right]
\end{align*}
is the \emph{channel dispersion}.
Neglecting the $o(1/\sqrt{\codelength})$ term in (\ref{eq:achieve-rate-coh}), we obtain the following approximation for $R_{\mathrm{coh}}^\ast(\codelength,\error)$
\begin{align}\label{eq:approx-coh}
R_{\mathrm{coh}}^\ast(\codelength,\error) \approx  C_{\mathrm{coh}}(\snr)-\sqrt{\frac{V_{\coh}(\snr)}{\codelength}}Q^{-1}(\error).
\end{align}
It was reported in~\cite{polyanskiy10-05,polyanskiy11-ge} that approximations similar to (\ref{eq:approx-coh}) are accurate for many channels for block-lengths and error probabilities of practical interest.
Hence, we will use~\eqref{eq:approx-coh} to evaluate $R_{\mathrm{coh}}^\ast(\codelength,\error)$ in the remainder of the paper.

\subsection{Upper Bound through Fano's inequality}
\label{sec:fano-upperbound}
\begin{figure}
  \centering
  \psfrag{UpperBound}[][][0.7]{$U(\snr)$ in (\ref{eq:capacit-upperbound})}
  \psfrag{LowerBound}[][][0.7]{$L(\snr)$ in (\ref{eq:lower-bound})}
  \psfrag{CohCapacity}[][][0.7]{$C_\coh(\snr)$ in (\ref{eq:capacity-coherent})}
  \psfrag{Xlabel}[][][0.7]{Channel's coherence time, $\cohtime$}
  \psfrag{Ylabel}[cB][rB][0.7]{Bits / channel use}
      \includegraphics[width=3.5in]{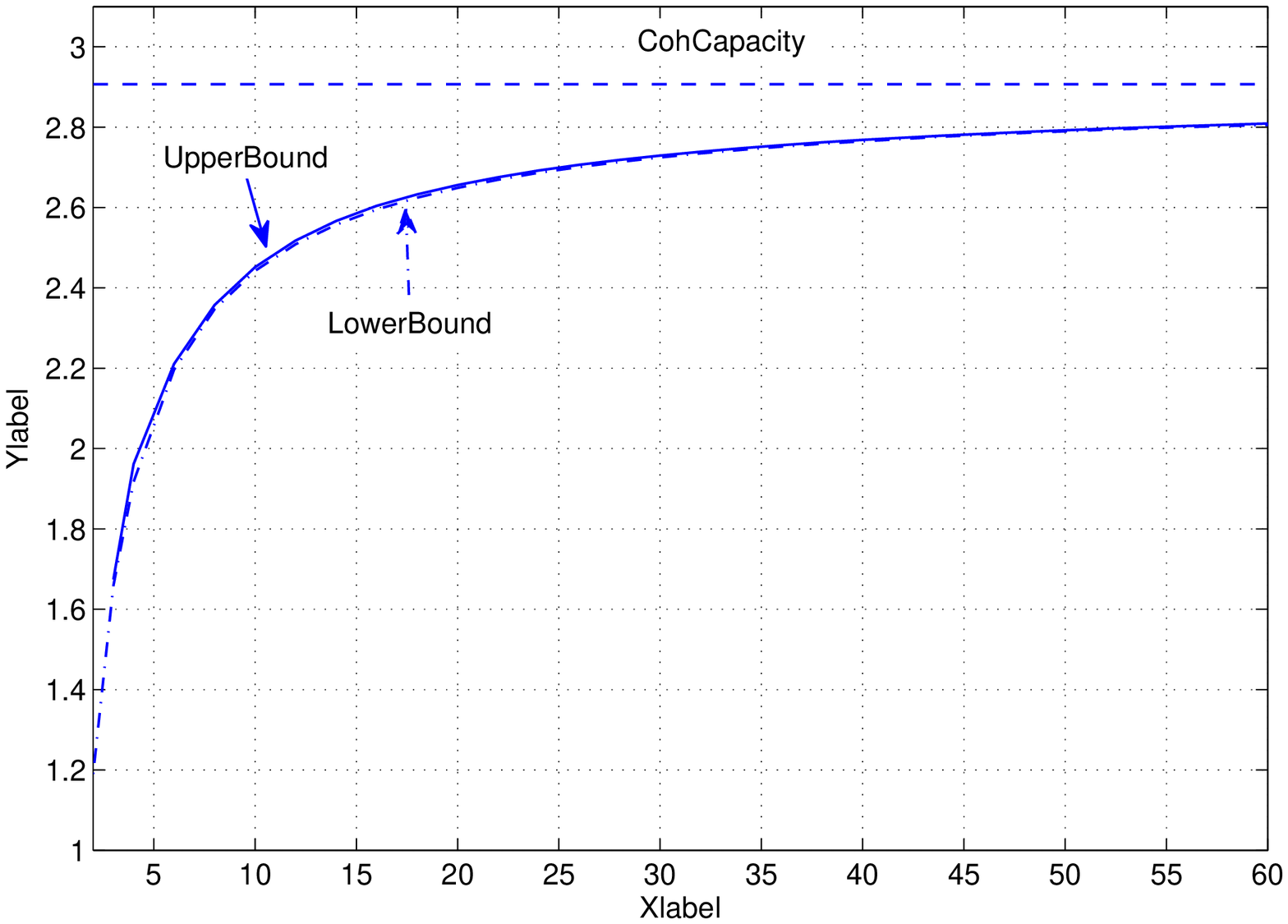}
      \caption{\label{fig:snr10db-capacity} $U(\snr)$ in (\ref{eq:capacit-upperbound}), $L(\snr)$ in (\ref{eq:lower-bound}) and $C_{\mathrm{coh}}(\snr)$ in (\ref{eq:capacity-coherent}) as a function of the channel's coherence time $\cohtime$, $\snr=10$ dB.}
\end{figure}
Our second upper bound follows from Fano's inequality~\cite[Thm. 2.10.1]{cover06-a}
\begin{align}
\label{eq:fano-inequality}
R^\ast(\codelength, \error ) \leq\frac{C(\snr) + H(\error)/\codelength}{1-\error}
\end{align}
where $H(x) = -x\log x -(1-x)\log(1-x)$ is the binary entropy function. Since no closed-form expression is available for $C(\snr)$, we will further upper-bound the right-hand side (RHS) of~\eqref{eq:fano-inequality} by replacing $C(\snr)$ with the capacity upper bound we shall derive below.

Let $\indist_{\vecy\given\vecx}$ denote the conditional distribution of $\vecy$ given $\vecx$, and $\indist_{\vecy}$ denote the distribution induced on $\vecy$ by the input distribution $\indist_{\vecx}$ through (\ref{eq:channel_IO_finitebl}). Furthermore, let $\outdist_{\vecy}$ be an arbitrary distribution on $\vecy$ with pdf $\outpdf_{\vecy}(\vecy)$. We can upper-bound $\mi(\vecx;\vecy)$ in (\ref{eq:def-capacity}) by duality as follows~\cite[Thm.~5.1]{lapidoth03-10a}:
\begin{align}\label{eq:def-duality}
\mi(\vecx;\vecy) 
&\leq \Ex{}{D(\indist_{\vecy\given\vecx}\|\outdist_{\vecy})}  \notag\\
&=-\Ex{\indist_{\vecy}}{\log\outpdf_{\vecy}(\vecy)}-\difent(\vecy\given \vecx).
\end{align}
Since
\begin{align}
\label{eq:average-power-constraint-2}
\cohtime\snr - \Ex{}{\|\vecx\|^2} \geq 0
\end{align}
for every $\indist_{\vecx}$ satisfying (\ref{eq:average-power-constraint}), we can upper bound $C(\snr)$ in (\ref{eq:def-capacity}) by using (\ref{eq:def-duality}) and (\ref{eq:average-power-constraint-2}) to obtain
\begin{IEEEeqnarray}{rcl}
\label{eq:capacity-ub-def}
C(\snr) \leq\frac{1}{\cohtime} \inf\limits_{\lambda\geq 0}\sup_{\indist_{\vecx}}\bigl\{&-&\Ex{\indist_{\vecy}}{\log\outpdf_{\vecy}(\vecy)}\notag\\
& - &\:\difent(\vecy\given \vecx)+\lambda(\cohtime\snr - \Ex{}{\|\vecx\|^2})\bigr\}.\IEEEeqnarraynumspace
\end{IEEEeqnarray}
The same bounding technique was previously used in \cite{katz04-10a} to obtain upper bounds on the capacity of the phase-noise AWGN channel (see also \cite{Martinez07}).

We next evaluate the RHS of (\ref{eq:capacity-ub-def}) for the following pdf
 \begin{equation}
\label{eq:outpdf-y}
\outpdf_\vecy(\vecy) = \frac{\Gamma(\cohtime)\|\vecy\|^{2(1-\cohtime)}}{\pi^\cohtime \cohtime (\snr+1)} e^{-\|\vecy\|^2/[\cohtime(\snr+1)]},\quad \vecy\in\complexset^{\cohtime}.
\end{equation}
Thus, $\vecy$ is i.d. and $\|\vecy\|^2 \sim \mathrm{Gamma}(1,\cohtime(1+\snr))$.
Substituting~(\ref{eq:outpdf-y}) into $\Ex{\indist_{\vecy}}{\log\outpdf_{\vecy}(\vecy)}$ in~\eqref{eq:capacity-ub-def}, we obtain
\begin{IEEEeqnarray}{rCl}
\IEEEeqnarraymulticol{3}{l}{-\Ex{\indist_{\vecy}}{\log \outpdf_\vecy(\vecy)}} \notag\\
\quad &=& \log\frac{\cohtime(1+\snr)\pi^{\cohtime}}{\Gamma(\cohtime)} + {\frac{\cohtime + \Ex{}{\|\vecx\|^2}} {\cohtime(\snr+1)}}\notag\\
&&+\:(\cohtime-1)\Ex{}{\log\lefto((1+\|\vecx\|^2)z_1+z_2 \right)}\notag\\
&=& \log\frac{\cohtime(1+\snr)\pi^{\cohtime}}{\Gamma(\cohtime)}+\frac{1}{\snr+1}+ (\cohtime-1)\psi(\cohtime-1)
\notag\\
&& +\: \Ex{}{(\cohtime-1)\sum\limits_{k=0}^{\infty}\frac{\left(1+1/\|\vecx\|^2\right)^{-k}}{k+\cohtime-1} + \frac{\|\vecx\|^2}{\cohtime(1+\snr)}}.\IEEEeqnarraynumspace\label{eq:exp-log-qy}
\end{IEEEeqnarray}
The first equality in (\ref{eq:exp-log-qy}) follows because the random variable $\|\vecy\|^2$ is conditionally distributed as $(1+\|\vecx\|^2)z_1+z_2$ given $\vecx$, where $z_1\sim\mathrm{Gamma}(1,1)$ and $z_2\sim\mathrm{Gamma}(\cohtime-1,1)$.

Substituting (\ref{eq:exp-log-qy}) into (\ref{eq:capacity-ub-def}), and using that the differential entropy $\difent(\vecy\given\vecx)$ is given by
\begin{align*}
\difent(\vecy\given\vecx) = \Ex{\vecx}{\log(1+\|\vecx\|^2)}+\cohtime\log(\pi e)
\end{align*}
we obtain
\begin{align}
\label{eq:capacit-upperbound-mid}
C(\snr)&\leq \frac{c_1}{\cohtime} + \frac{1}{\cohtime} \inf_{\lambda\geq 0}\sup_{\indist_{\vecx}}\left\{ \mathbb{E}\lefto[  \sum\limits_{k=0}^{\infty}\frac{(\cohtime-1)\left(1+1/\|\vecx\|^2\right)^{-k}}{k+\cohtime-1} \right.\right.\notag\\
&\quad\left.\left.\vphantom{\sum\limits_{k=0}^{\infty}\frac{(\cohtime-1)\left(1+1/\|\vecx\|^2\right)^{-k}}{k+\cohtime-1}}-\log\lefto(1+\|\vecx\|^2\right)+ \frac{\|\vecx\|^2}{\cohtime(1+\snr)}  +\lambda\lefto(\cohtime \snr-\|\vecx\|^2\right) \right]\right\}\\
&\stackrel{(a)}{\leq} \frac{c_1}{\cohtime} + \frac{1}{\cohtime} \inf_{\lambda\geq 0}\sup_{\|\vecx\|}\left\{  \sum\limits_{k=0}^{\infty}\frac{(\cohtime-1)\left(1+1/\|\vecx\|^2\right)^{-k}}{k+\cohtime-1}\right.\notag\\
&\quad \left.\vphantom{\sum\limits_{k=0}^{\infty}\frac{(\cohtime-1)\left(1+1/\|\vecx\|^2\right)^{-k}}{k+\cohtime-1}}-\log\lefto(1+\|\vecx\|^2\right) + \frac{\|\vecx\|^2}{\cohtime(1+\snr)}  +\lambda\left(\cohtime\snr-\|\vecx\|^2\right) \right\}\notag\\
\label{eq:capacit-upperbound}&\define U(\snr)
\end{align}
where
\begin{align*}
c_1\define\log\frac{\cohtime(1+\snr)}{\Gamma(\cohtime)}-\cohtime +\frac{1}{\snr+1} + (\cohtime -1)\psi(\cohtime-1).
\end{align*}
To obtain (a), we upper-bounded the second term on the RHS of (\ref{eq:capacit-upperbound-mid}) by replacing the expectation over $\|\vecx\|$ by the supremum over $\|\vecx\|$.

The bounds $L(\snr)$ and $U(\snr)$ are plotted in Fig.~\ref{fig:snr10db-capacity} as a function of the channel's coherence time $\cohtime$ for SNR equal to 10 dB. For reference, we also plot the capacity in the coherent setting [$C_{\mathrm{coh}}(\snr)$ in (\ref{eq:capacity-coherent})]. We observe that $U(\snr)$ and $L(\snr)$ are surprisingly close for all values of \cohtime.

At low SNR, the gap between $U(\snr)$ and $L(\snr)$ increases. In this regime, $U(\snr)$ can be tightened by replacing $\outpdf_{\vecy}(\vecy)$ in (\ref{eq:capacity-ub-def}) by the output pdf induced by the i.d. unitary input distribution $\indist^{(\mathrm{U})}_{\vecx}$, which is given by
\begin{multline}
\label{eq:outpdf-induced}
\outpdf^{(\mathrm{U})}_\vecy(\vecy)
= \frac{e^{-\|\vecy\|^2/(1+\cohtime\snr)} \|\vecy\|^{2(1-\cohtime)}\Gamma(\cohtime)}
{{\pi^\cohtime}(1+\cohtime\snr) } \\
\quad\quad\times \tilde{\gamma}\lefto(\cohtime-1, \frac{\cohtime\snr \|\vecy\|^2}{1+\cohtime\snr}\right)\left(1+\frac{1}{\cohtime\snr}\right)^{\cohtime-1}.
\end{multline}

Substituting (\ref{eq:capacit-upperbound}) into (\ref{eq:fano-inequality}), we obtain the following upper bound on $R^\ast(\codelength,\error)$:
\begin{align}
\label{eq:upperbound-fano}
R^\ast(\codelength,\error) \le \bar{R}(\codelength,\error) \define \frac{U(\snr) + H(\error)/\codelength}{1-\error}.
\end{align}

\subsection{Dependence Testing (DT) Lower Bound}\label{sec:dt-lb}
We next present a lower bound on $R^\ast(\codelength, \error)$ that is based on the DT bound recently proposed by Polyanskiy, Poor, and Verd\'{u}~\cite{polyanskiy10-05}.
The DT bound uses a threshold decoder that sequentially tests all messages and returns the first message whose likelihood exceeds a pre-determined threshold. With this approach, one can show that for a given input distribution $\indist_{\vecx^\blocklength}$, there exists a code with $M$ codewords and average probability of error not exceeding~\cite[Thm.~17]{polyanskiy10-05}
\begin{align}
\label{eq:DT-bound-1}
\epsilon &\leq \mathbb{E}_{\indist_{\vecx^\blocklength}}\lefto[\indist_{\vecy^\blocklength \given\vecx^\blocklength}\lefto(i\lefto(\vecx^\blocklength;\vecy^\blocklength\right)\leq \log\frac{\NumCode-1}{2}\right)\right.\notag\\
&\qquad\quad\quad + \left.\frac{\NumCode-1}{2} \indist_{\vecy^\blocklength}\lefto(i\lefto(\vecx^\blocklength;\vecy^\blocklength\right)> \log\frac{\NumCode-1}{2}\right)\right]
\end{align}
where
\begin{equation}
\label{eq:information-density}
i\lefto(\vecx^\blocklength;\vecy^\blocklength\right) \define \log\frac{\inpdf_{\vecy^\blocklength \given \vecx^\blocklength}\lefto(\vecy^\blocklength\given \vecx^\blocklength\right)}{\inpdf_{\vecy^\blocklength}\lefto(\vecy^\blocklength\right)}
\end{equation}
is the \emph{information density}. Note that, conditioned on $\vecx^\blocklength$, the output vectors $\vecy_l$, $l=1,\ldots,L$, are independent and Gaussian distributed. The pdf of $\vecy_l$ is given by
\begin{align}
\label{eq:conditional-pdf}
&\inpdf_{\vecy\given \vecx}(\vecy_l\given \vecx_l) \notag\\
&\quad = \frac{\exp\lefto(-\herm{\vecy_l}(\matI_\cohtime + \vecx_l\herm{\vecx_l})^{-1}\vecy_l\right)}{\pi^\cohtime\det(\matI_\cohtime + \vecx_l\herm{\vecx_l})}\notag\\
&\quad\stackrel{(a)}{=}\frac{1}{\pi^\cohtime(1+\|\vecx_l\|^2)}\exp\lefto(-\|\vecy_l\|^2 +  \frac{|\herm{\vecy_l}\vecx_l|^2}{1+\|\vecx_l\|^2} \right)
\end{align}
where (a) follows from Woodbury's matrix identity \cite[p.~19]{horn85a}.

To evaluate (\ref{eq:DT-bound-1}), we choose $\vecx_l$, $l=1,\ldots,\blocklength$, to be independently and identically distributed according to the i.d. unitary distribution $\indist_\vecx^{(\mathrm{U})}$. The pdf of the corresponding output distribution is equal to
\begin{equation*}
\label{eq:marginal-pdf}
\outpdf^{(\mathrm{U})}_{\vecy^\blocklength }(\vecy^\blocklength) = \prod \limits^{\blocklength}_{l=1} \outpdf_\vecy^{(\mathrm{U})}(\vecy_l)
\end{equation*}
where $\outpdf_\vecy^{(\mathrm{U})}(\cdot)$ is given in (\ref{eq:outpdf-induced}).
Substituting (\ref{eq:conditional-pdf}) and (\ref{eq:outpdf-induced}) into (\ref{eq:information-density}), we obtain
\begin{equation}
\label{eq:dt_infden-L}
 i\lefto(\vecx^\blocklength; \vecy^\blocklength\right) = \sum\limits_{l=1}^{\blocklength} i(\vecx_l;\vecy_l)
 \end{equation}
 where
 \begin{align*}
 i(\vecx_l;\vecy_l)
 & = \log \frac{1+\cohtime\snr}{\Gamma(\cohtime)} + \frac{|\herm{\vecy_l}\vecx_l|^2}{1+\|\vecx_l\|^2} - \frac{\cohtime \snr \|\vecy_l\|^2}{1+\cohtime \snr}  \notag\\
 &\quad +(\cohtime -1)\log\frac{\cohtime\snr  \|\vecy_l\|^2}{1+\cohtime\snr} - \log\lefto(1+\|\vecx_l\|^2\right)\notag\\
 & \quad -   \log \tilde{\gamma}\lefto(\cohtime-1, \frac{\cohtime\snr \|\vecy_l\|^2}{1+\cohtime\snr}\right).
 \end{align*}

Due to the isotropy of both the input distribution $\indist^{(\mathrm{U})}_{\vecx^\blocklength}$ and the output distribution $\outdist^{(\mathrm{U})}_{\vecy^\blocklength}$, the distribution of the information density $i\lefto(\vecx^\blocklength;\vecy^\blocklength\right)$ depends on $\indist^{(\mathrm{U})}_{\vecx^\blocklength}$ only through the distribution of the norm of the inputs $\vecx_l$. Furthermore, under $\indist^{(\mathrm{U})}_{\vecx^\blocklength}$, we have that $\|\vecx_l\|=\sqrt{\cohtime\snr}$ with probability 1, $l=1,\ldots,\blocklength$. This allows us to simplify the computation of (\ref{eq:DT-bound-1}) by choosing an arbitrary input sequence
$\vecx_l= \bar{\vecx}\define\tp{[\sqrt{\cohtime\snr},0,\ldots,0]}$, $l=1,\ldots,\blocklength$.
Substituting (\ref{eq:dt_infden-L}) into (\ref{eq:DT-bound-1}), we obtain the desired lower bound on $R^\ast(\codelength,\error)$ by solving numerically the following maximization problem
\begin{align}
\label{eq:DT-bound-2}
\underline{R}(\codelength, \error)\define \max \left\{\frac{1}{\codelength}\log \NumCode : \NumCode\text{ satisfies (\ref{eq:DT-bound-1})} \right\}.
\end{align}
\begin{figure}
  \centering
   \psfrag{UpperBound}[][][0.7]{$\bar{R}(\codelength,\error)$ in (\ref{eq:upperbound-fano})}
  \psfrag{DTbound}[][][0.7]{$\underline{R}(\codelength,\error)$ in (\ref{eq:DT-bound-2})}
  \psfrag{Secondline}[][][0.7]{$\&$ approximation (\ref{eq:approx-noncoh})}
  \psfrag{CohCapacity}[][][0.7]{$C_\coh(\snr)$ in (\ref{eq:capacity-coherent})}
  \psfrag{CohApprox}[][][0.7]{Approximation of $R^\ast_{\coh}(\codelength,\error)$ in (\ref{eq:approx-coh})}
  \psfrag{Xlabel}[][][0.7]{Block-length, $\codelength$}
  \psfrag{Ylabel}[cB][rB][0.7]{Bits / channel use}
      \includegraphics[width=3.5in]{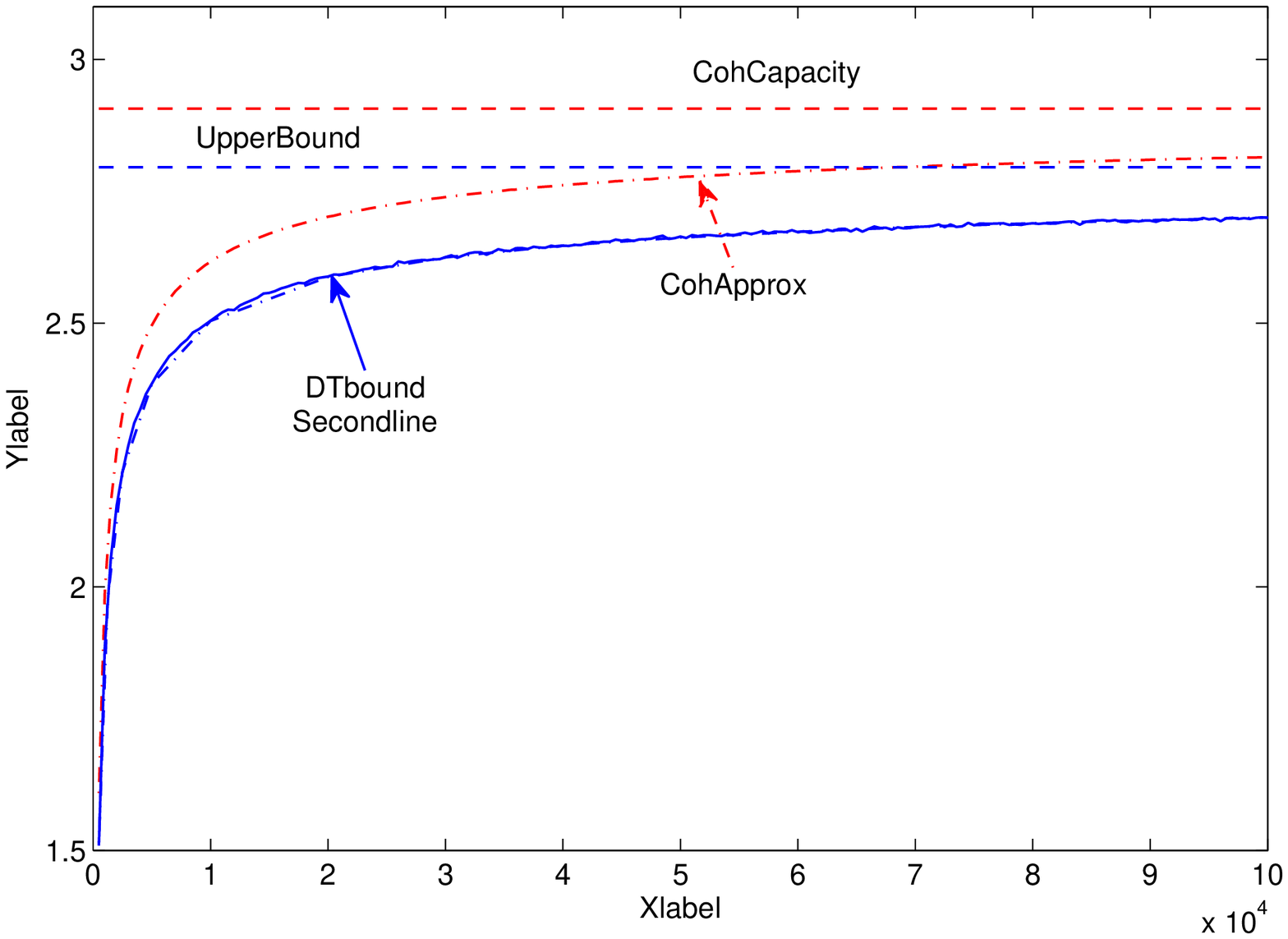}
      \caption{\label{fig:finite-bl-bound} Bounds on maximal achievable rate $R^\ast(\codelength,\error)$ for noncoherent Rayleigh block-fading channels; $\snr=10$ dB, $\cohtime=50$, $\error=10^{-3}$.}
\end{figure}

The computation of the DT bound $\underline{R}(\codelength, \error)$ becomes difficult as the block-length $\codelength$ becomes large. We next provide an approximation for $\underline{R}(\codelength, \error)$, which is much easier to evaluate. As in \cite[App. A]{polyanskiy11-ge}, applying \emph{Berry-Esseen inequality}\cite[Thm.~44]{polyanskiy10-05} to the first term on the RHS of (\ref{eq:DT-bound-1}), and applying \cite[Lemma 20]{Polyanskiy10} to the second term on the RHS of (\ref{eq:DT-bound-1}), we get the following asymptotic expansion for $\underline{R}(\codelength, \error)$
\begin{align}
\label{eq:lb-expansion-noncoherent}
\underline{R}(\codelength, \error) = L(\snr) - \sqrt{\frac{ \underline{V}(\snr)}{\codelength}}Q^{-1}(\error) + \bigo\lefto(\frac{1}{\codelength}\right), \codelength \rightarrow \infty
\end{align}
with $\underline{V}(\snr)$ given by
\begin{align*}
\underline{V}(\snr) &\define \frac{1}{\cohtime}\Ex{\indist_{\vecx}^{(\mathrm{U})}}{\mathrm{Var}\lefto[i(\vecx;\vecy)\given \vecx\right]}=\frac{1}{\cohtime}\mathrm{Var}\lefto[i(\bar{\vecx};\vecy)\right]
\end{align*}
where, as in the DT bound, we can choose $\bar{\vecx}=\tp{[\sqrt{\cohtime\snr},0,\ldots,0]}$.
By neglecting the $\bigo(1/\codelength)$ term in (\ref{eq:lb-expansion-noncoherent}), we arrive at the following approximation for $\underline{R}(\codelength, \error)$
\begin{align}
\label{eq:approx-noncoh}
\underline{R}(\codelength, \error) \approx L(\snr) - \sqrt{\frac{ \underline{V}(\snr)}{\codelength}}Q^{-1}(\error).
\end{align}
Although the term $\underline{V}(\snr)$ in (\ref{eq:approx-noncoh}) needs to be computed numerically, the computational complexity of (\ref{eq:approx-noncoh}) is much lower than that of the DT bound $\underline{R}(\codelength, \error)$.

\subsection{Numerical Results and Discussions}

In Fig.~\ref{fig:finite-bl-bound}, we plot the upper bound $\bar{R}(\codelength, \error)$ in (\ref{eq:upperbound-fano}), the lower bound $\underline{R}(\codelength,\error)$ in (\ref{eq:DT-bound-2}), the approximation of $\underline{R}(\codelength,\error)$ in (\ref{eq:approx-noncoh}), and the approximation of $R^\ast_{\mathrm{coh}}(\codelength, \error)$ in (\ref{eq:approx-coh}) as a function of the block-length $\codelength$ for $\cohtime=50$, $\error=10^{-3}$ and $\snr=10$ dB.
For reference, we also plot the coherent capacity $C_{\mathrm{coh}}(\snr)$ in (\ref{eq:capacity-coherent}).
As illustrated in the figure, (\ref{eq:approx-noncoh}) gives an accurate approximation of $\underline{R}(\codelength,\error)$.
\begin{figure}
  \centering
    \psfrag{UpperBound}[][][0.7]{$\bar{R}(\codelength,\error)$ in (\ref{eq:upperbound-fano})}
  \psfrag{DTbound}[][][0.7]{$\underline{R}(\codelength,\error)$ in (\ref{eq:DT-bound-2})}
  \psfrag{CohCapacity}[][][0.7]{$C_\coh(\snr)$ in (\ref{eq:capacity-coherent})}
  \psfrag{CohApprox}[][][0.7]{Approximation of $R^\ast_{\coh}(\codelength,\error)$ in (\ref{eq:approx-coh})}
  \psfrag{Xlabel}[][][0.7]{Channel's coherence time, $\cohtime$}
  \psfrag{Ylabel}[cB][rB][0.7]{Bits / channel use}
  \psfrag{BlockLength}[][][0.7]{$\codelength=4\times10^3$}
  \psfrag{Error}[cB][rB][0.7]{$\error=10^{-3}$}
      \includegraphics[width=3.5in]{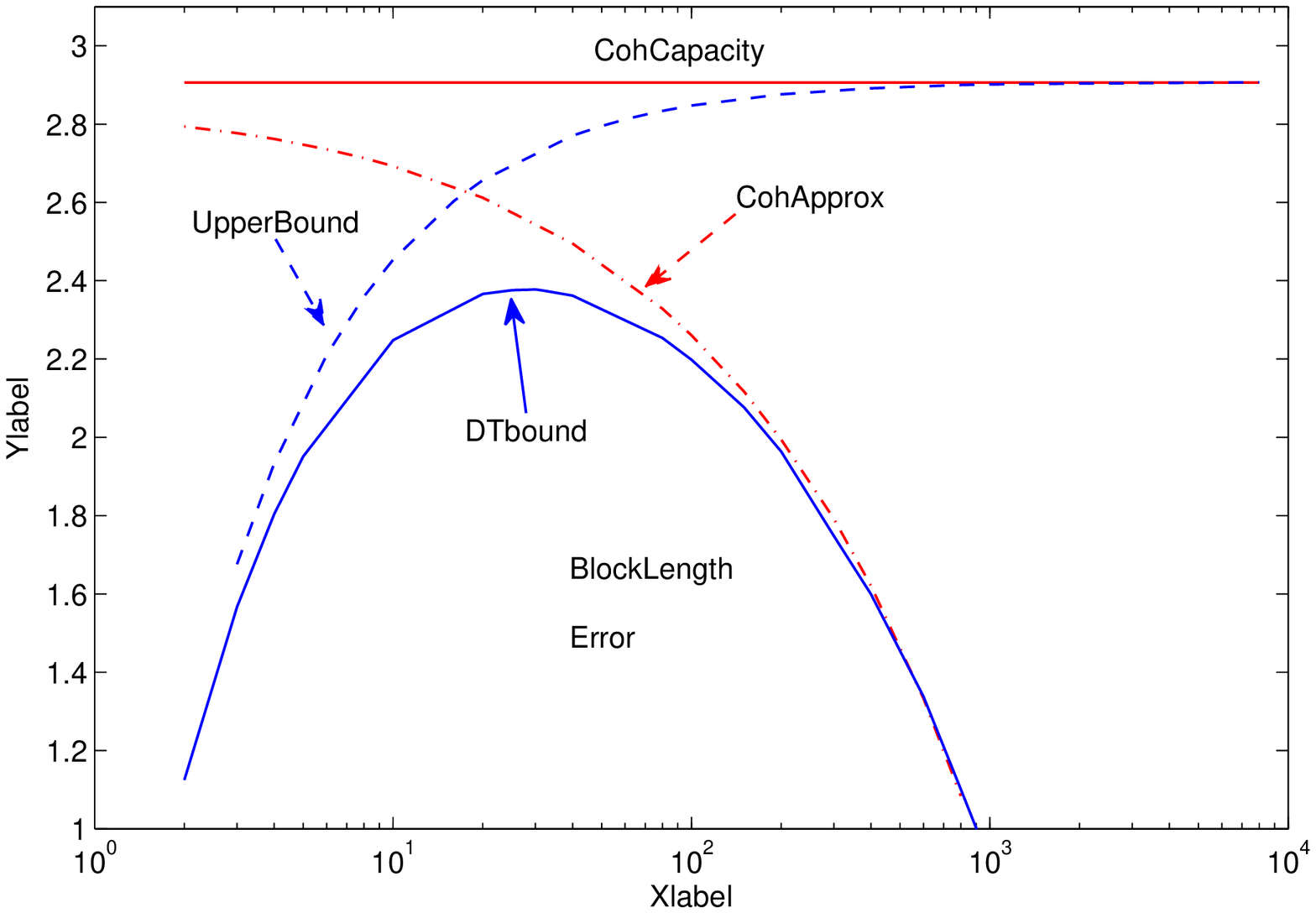}
      \caption{\label{fig:fix_bl_4000} $\bar{R}(\codelength,\error)$ in (\ref{eq:upperbound-fano}), $\underline{R}(\codelength,\error)$ in (\ref{eq:DT-bound-2}), approximation of $R^\ast_{\coh}(\codelength,\error)$ in (\ref{eq:approx-coh}), and $C_\coh(\snr)$ in (\ref{eq:capacity-coherent}) at block-length $\codelength = 4\times 10^{3}$ as a function of the channel's coherence time $\cohtime$ for the noncoherent Rayleigh block-fading channel; $\snr=10$ dB, $\epsilon=10^{-3}$.}
      \end{figure}
\begin{figure}
  \centering
     \psfrag{UpperBound}[][][0.7]{$\bar{R}(\codelength,\error)$ in (\ref{eq:upperbound-fano})}
  \psfrag{DTbound}[][][0.7]{$\underline{R}(\codelength,\error)$ in (\ref{eq:DT-bound-2})}
  \psfrag{CohCapacity}[][][0.7]{$C_\coh(\snr)$ in (\ref{eq:capacity-coherent})}
  \psfrag{CohApprox}[][][0.7]{Approximation of $R^\ast_{\coh}(\codelength,\error)$ in (\ref{eq:approx-coh})}
  \psfrag{Xlabel}[][][0.7]{Channel's coherence time, $\cohtime$}
  \psfrag{Ylabel}[cB][rB][0.7]{Bits / channel use}
  \psfrag{BlockLength}[][][0.7]{$\codelength=4\times10^4$}
  \psfrag{Error}[cB][rB][0.7]{$\error=10^{-3}$}
      \includegraphics[width=3.5in]{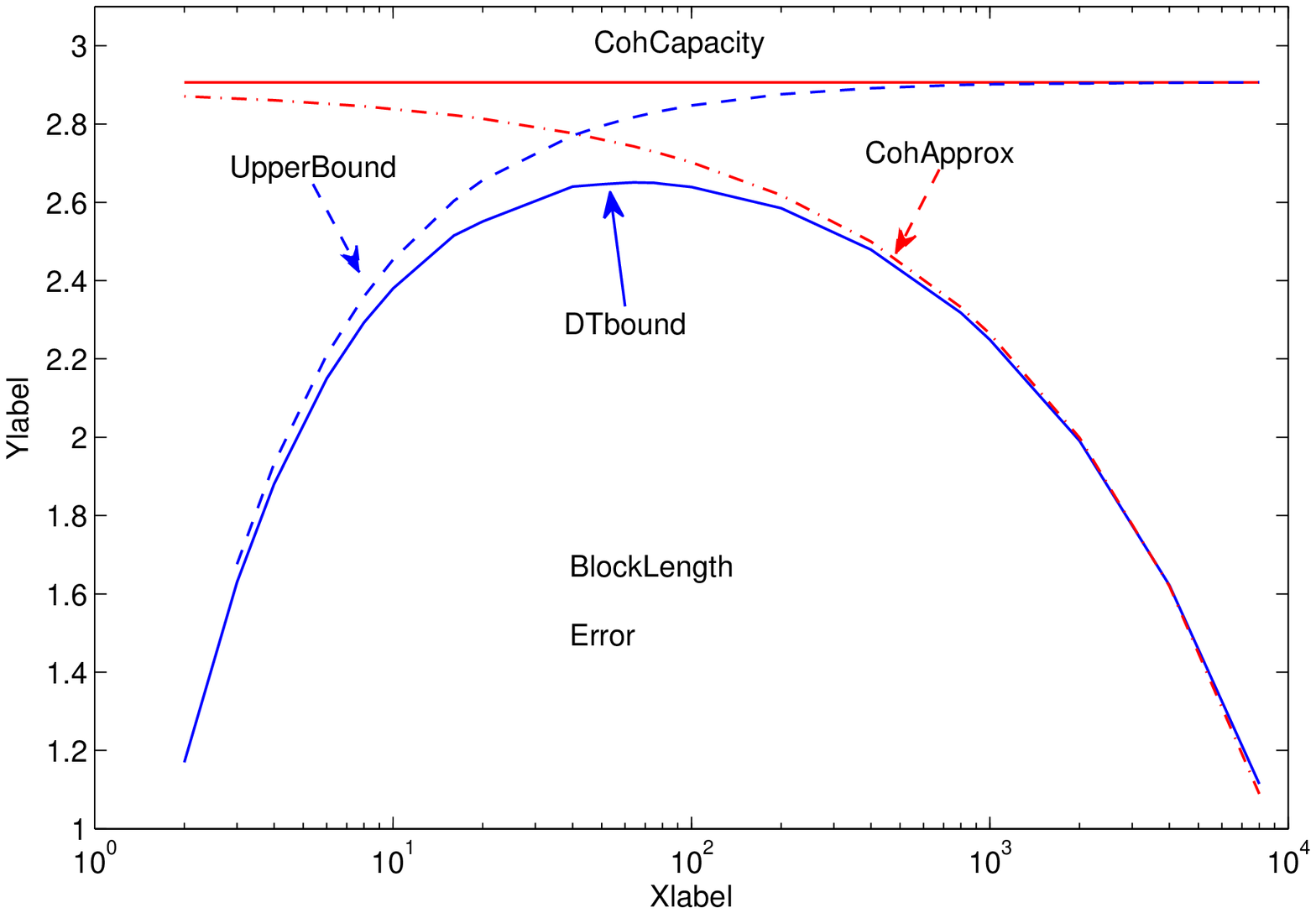}
      \caption{\label{fig:fix_bl_40000} $\bar{R}(\codelength,\error)$ in (\ref{eq:upperbound-fano}), $\underline{R}(\codelength,\error)$ in (\ref{eq:DT-bound-2}), approximation of $R^\ast_{\coh}(\codelength,\error)$ in (\ref{eq:approx-coh}), and $C_\coh(\snr)$ in (\ref{eq:capacity-coherent}) at block-length $\codelength = 4\times 10^{4}$ as a function of the channel's coherence time $\cohtime$ for the noncoherent Rayleigh block-fading channel; $\snr=10$ dB, $\epsilon=10^{-3}$.}
      \end{figure}

In Figs.~\ref{fig:fix_bl_4000} and \ref{fig:fix_bl_40000}, we plot the upper bound $\bar{R}(\codelength, \error)$ in (\ref{eq:upperbound-fano}), the lower bound $\underline{R}(\codelength,\error)$ in (\ref{eq:DT-bound-2}), the approximation of $R^\ast_{\mathrm{coh}}(\codelength, \error)$ in (\ref{eq:approx-coh}), and the coherent capacity $C_{\coh}(\snr)$ in (\ref{eq:capacity-coherent}) as a function of the channel's coherence time $\cohtime$ for block-lengths $\codelength = 4\times10^3$ and $\codelength = 4\times 10^4$, respectively. We see that, for a given block-length and error probability, $R^\ast(\codelength,\error)$ is not monotonic in the channel's coherence time, but there exists a channel's coherence time $\cohtime^\ast$ that maximizes
$R^\ast(\codelength,\error)$. This confirms the claim we made in the introduction that there exists a tradeoff between the diversity gain and the cost of estimating the channel when communicating in the noncoherent setting and in the finite block-length regime. A similar phenomenon was observed in \cite{polyanskiy11-ge} for the Gilbert-Elliott channel with no state information at the transmitter and receiver.
\balance

From Figs.~\ref{fig:fix_bl_4000} and \ref{fig:fix_bl_40000}, we also observe that $\cohtime^\ast$ decreases as we shorten the block-length. For example, the rate-maximizing channel's coherence time $\cohtime^\ast$ for block-length $\codelength=4\times10^4$ is roughly 64, whereas for block-length $\codelength = 4\times10^3$, it is roughly $28$.


\bibliographystyle{IEEEtran}
\bibliography{IEEEabrv,publishers,confs-jrnls,WeiBib}

\end{document}